\documentclass{aastex}
\usepackage{spr-astr-addons}
\usepackage{url}

\RequirePackage{color}

\usepackage[colorlinks=true,
            linkcolor=red,
            urlcolor=blue,
            citecolor=cyan]{hyperref}

\begin{document}

\title{Hawking radiation of Dirac monopoles from the global monopole black hole with quantum gravity effects}
\slugcomment{Not to appear in Nonlearned J., 45.}
\shorttitle{Short article title}
\shortauthors{Autors et al.}

\author{Kimet Jusufi\altaffilmark{1}} \and \author{Gordana Apostolovska \altaffilmark{2}}

\altaffiltext{1}{Physics Department, State University of Tetovo, Ilinden Street nn, 1200,
Macedonia, \url{kimet.jusufi@unite.edu.mk}}
\altaffiltext{2}{Institute of Physics, Faculty of Natural Sciences and Mathematics,
Ss. Cyril and Methodius University, Arhimedova 3, 1000 Skopje, Macedonia, \url{gordanaa@pmf.ukim.mk}}

\begin{abstract}
In this paper we study the quantum tunneling of Dirac magnetic monopoles from the global monopole black hole under quantum gravity effects. We start from the modified Maxwell's equations and the Generalized Uncertainty Relation (GUP), to recover the GUP corrected temperature for the global monopole black hole by solving the modified Dirac equation via Hamilton-Jacobi method.  Furthermore, we also include the quantum corrections beyond the semiclassical approximation, in particular, first we find the logarithmic corrections of GUP corrected entropy and finally we calculate the GUP corrected specific heat capacity. It is argued that the GUP effects may prevent a black hole from complete evaporation and leave remnants. 
\end{abstract}

\keywords{Hawking radiation, Generalized uncertainty relation, Magnetic monopoles,
Global monopoles}


\section{Introduction}

The discovery of Hawking radiation \cite{Hawking1}, has opened up new exploration of research which combines quantum mechanics, general relativity, and thermodynamics. Hawking radiation has also profound implications, namely, the black hole evaporation process results with the information loss paradox. Today, however, the situation has changed, this radiation continues not only to attract attention of pure theoretical reasons, but there are good reason to believe that, this phenomen could be experimentally observed by using an analogue black hole \cite{Steinhauer}.

Hawking radiation has been studied extensively in the past and shown that one can recover the same black hole temperature to a particular black hole configuration in different methods.  For example, the Euclidean path integral approach of Gibbons and Hawking, tunneling method, dimensional reduction near the horizon  \cite{gibbons1,gibbons2,umetso,kraus1,kraus2,perkih1,perkih2,perkih3,ang,sri1,sri2,vanzo}. In the original method used by Hawking, this radiation is shown to be thermal, therefore one wonders if this is the case also for the other methods. The tunneling method shows that this is not true, in fact, one can apply this method for different kind of particles from various types of black holes and show that the radiation spectrum can deviate from pure thermality \cite{mann0,mann1,mann2,mann3,sakalli1,sakalli2,kruglov1,xiang,gohar1,ren,kimet,ahmed1}. This, may, after all suggest that the information is preserved, but unfortunately doesn't solve the information loss paradox.  On the other hand, Banerjee and Majhi \cite{banerjee,akbar,majumder} investigated the modified temperature and modified entropy  beyond the semiclassical approximation. Recently, number of authors \cite{qgv1,qgv2,ali,ali1,brito1,brito2,khalil,faizal1,faizal2}, incorporated the effects of quantum gravity for scalar particles and fermions to show that GUP effects slows down the increase of the Hawking temperature. 

The existence of magnetic monopoles currently is ruled out by experiments. However, Maxwell's equations can be modified and written in the presence of electric and magnetic sources. Furthermore, Dirac showed that, if magnetic monopoles exist, their charge must come in quanta of a certain size \cite{dirac}, also known as Dirac monopoles. In this context, Hawking radiation has also been investigated for charged and magnetized particles \cite{li1,li2,btz,zeng2,zhang}. It is widely believed that monopoles may have been produced in the early universe and then diluted away during the inflation. The theory behind magnetic monopoles attracted interest again when 't Hooft and Polyakov \cite{thooft,polyakov} in 1974 argued about the existence of a magnetic monopole which is fundamentally different from the Dirac magnetic monopole. Global monopoles on the other hand are heavy objects that may be produced during the phase transitions in the early universe \cite{vilenkin}. Many authors have speculated a possible connection between magnetic monopoles and global monopoles, for example in Ref. \cite{spinelly}, a regular global monopole with a magnetic field is described. Such scenario, however,  might have astrophysical implications  manifested by the magnetic field if such objects exists, say, in a typical galaxy. In Ref. \cite{yu}, the massive scalar particles was studied in the background of a black hole with a global monopole. More recently, tunneling beyond the semiclassical approximation with global monopole and other methods \cite{zeng1,qing}, wheares in Ref. \cite{chen}, the GUP effect was investigated in the context of $f(R)$ global monopole. Since global monopoles may be supermassive and carry magnetic charges, it is natural, therefore, to study the quantum tunneling of magnetic monopoles (Dirac monopoles) from the global monopole black hole. Morover, we will also investigate the quantum gravity effects on the Hawking radiation of magnetic monopoles by calculating the GUP temperature and GUP entropy beyond the semiclassical approximation. To the best of our knowledge, the tunneling of magnetic monopoles from global monopole black hole under GUP effects has not been yet investigated in the literature.

The paper is organized as follows. In Sec. II, we review the Maxwell's equations with electric and magnetic charges. In Sec. III, we solve the modified Dirac equation for charged/magnetized fermions and calculate the Hawking temperature under GUP effects from a global monopole black hole. In Sec. IV, we explore the GUP effects on entropy and specific heat capacity of the black hole beyond the semiclassical approximation. In Sec. V, we comment on our results.

\section{Modified Maxwell's Equations with Magnetic
Charges}

The standard Maxwell's equations forbid the existence of magnetic monopoles. Dirac on the other hand, first predicted the existence of the magnetic monopole. The electromagnetic tensor with a source with electric and magnetic charge is written as \cite{zhang}
\begin{equation}
F_{\mu\nu}=\nabla_{\nu}A_{\mu}-\nabla{_\mu}A_{\nu}+G_{\mu\nu}^{+}\label{1}
\end{equation}
where $G_{\mu\nu}^{+} $ is the Dirac string term. Furthermore the Maxwell equations can be written as

\begin{equation}
\nabla_{\nu}F^{	\mu\nu}=4 \pi J^{\mu}_{e},\label{2}
\end{equation}
\begin{equation}
\nabla_{\nu}F^{+\mu\nu}=4\pi J^{\mu}_{m},\label{3}
\end{equation}
in which $F^{+\mu\nu}$ is the dual tensor of $F^{\mu\nu}$, while $J^{\mu}_{m}=\rho_{m}u^{\mu}$ and $J^{\mu}_{e}=\rho_{e}u^{\mu}$ are the four electric and magnetic currents, respectively. Morover, $\rho_{e}$ and $\rho_{m}$, represent the densities of electric and
magnetic charges, while $u^{\mu}$ stand for the 4--velocity. It is convenient to introduce a new real anti-symmetric tensor defined as \cite{li1,li2}
\begin{equation}
\tilde{F}^{\mu\nu}=F^{\mu\nu}\cos\beta+F^{+\mu\nu}\sin\beta,\label{4}
\end{equation}
in which $\beta$ denotes a real constant angle. By substituting this equation into the Eqs. \eqref{2} and  \eqref{3} it follows that
\begin{equation}
\nabla_{\nu}\tilde{F}^{\mu\nu}=4\pi \left(\rho_{e}\cos\beta+\rho_{m}\sin\beta\right)u^{\mu},\label{5}
\end{equation}
\begin{equation}
\nabla_{\nu}\tilde{F}^{+\mu\nu}=4\pi \left(-\rho_{e}\sin\beta+\rho_{m}\cos\beta\right)u^{\mu}.\label{6}
\end{equation}

Now, the above equations can be simplified by letting
\begin{equation}
\rho_{e}\cos\beta+\rho_{m}\sin\beta=\rho_{h},\label{7}
\end{equation}
\begin{equation}
-\rho_{e}\sin\beta+\rho_{m}\cos\beta=0, \label{8}
\end{equation}
which suggests that, one can now recover an analogue form to the Maxwell's equations as follows
\begin{equation}
\nabla_{\nu}\tilde{F}^{\mu\nu}=4\pi \rho_{h}u^{\mu},\label{9}
\end{equation}
\begin{equation}
\nabla_{\nu}\tilde{F}^{+\mu\nu}=0.\label{10}
\end{equation}

In which the corresponding electromagnetic tensor reads $\tilde{F}_{\mu\nu}=\nabla_{\nu}\tilde{A}_{\mu}-\nabla_{\mu}\tilde{A}_{\nu}$. If we introduce the current densities  $J^{\mu}=\rho_{h}u^{\mu}$, then Eq. \eqref{9}, can also be written as
\begin{equation}
\frac{\partial( \sqrt{-g} \tilde{F}^{\mu\nu})}{\partial x^{\nu}}=4\pi \sqrt{-g}J^{\mu}.\label{11}
\end{equation}

In other words, a charged black hole, can be characterized by the equivalent charge $Q_{h}$, and equivalent charge density $\rho_{h}$. The electric charge and the magnetic charge are concentrated on the black hole with the density rate given as $\rho_{e}/\rho_{m}=\cos\beta$, so that
\begin{equation}
Q_{h}^{2}=Q_{e}^{2}+Q_{m}^{2}.\label{12}
\end{equation}

Finally, the Lagrangian density of the electromagnetic field can be given as follows
\begin{equation}
\mathcal{L}_{h}=-\frac{1}{4}\tilde{F}_{\mu\nu}\tilde{F}^{\mu\nu}
\end{equation}

In the next section, we will study the tunneling of charged and magnetized fermions from the charged global monopole black hole. In particular, first we will analyze the effects of GUP on the Hawking radiation, and then we will also analyze the GUP effects on entropy and specific heat capacity of the black hole.

\section{Tunneling From Global Monopole Black Hole with GUP Effects}

The metric of the Reissner--Nordstr\"{o}m black hole with magnetic charges in the spacetime with a global monopole can be written as \cite{vilenkin}
\begin{equation}
ds^{2}=-g(\tilde{r})d{\tilde{t}}^{2}+g({\tilde{r}})^{-1}d{\tilde{r}}^{2}+{\tilde{r}}^{2}\left(d\theta^{2}+\sin^{2}\theta d\varphi^{2} \right),\label{13}
\end{equation}
in which
\begin{equation}
g({\tilde{r}})=\left(1-8\pi\eta^{2}-\frac{2\tilde{M}}{\tilde{r}}+\frac{\tilde{Q}^{2}_{h}}{{\tilde{r}}^{2}}\right).\label{14}
\end{equation}

One such an interesting scenario might happens if a charged black hole has swallowed a global monopole carrying magnetic charges. Note that here $M$ is the black hole mass parameter, $Q_{h}$ gives the equivalent charge parameter of the black hole, and $\eta$ is the symmetry
breaking scale when the monopole is formed. This metric can be symplified by introducing the following coordinate transformations \cite{yu}
\begin{equation}
\tilde{t}\to \left(1-8\pi\eta^{2}\right)^{-1/2}t, \,\,\,\,\, \tilde{r}\to \left(1-8\pi\eta^{2}\right)^{1/2}r , \label{15}
\end{equation}
and by defining new parameters as
\begin{equation}
\tilde{M}  \to  \left(1-8\pi \eta^{2}\right)^{3/2}M,\,\,\,\,\,\tilde{Q}_{h}\to Q_{h} \left(1-8\pi\eta^{2}\right). \label{16}
\end{equation}

The metric \eqref{13}, now takes the following form
\begin{equation}
ds^{2}=-f(r)dt^{2}+f(r)^{-1}dr^{2}+a\, r^{2}\left(d\theta^{2}+\sin^{2}\theta d\varphi^{2} \right), \label{17}
\end{equation}
in which $a=(1-8\pi\eta^{2})$, and 
\begin{equation}
f(r)=1-\frac{2M}{r}+\frac{Q_{h}^{2}}{r^{2}}=\frac{(r-r_{+})(r-r_{-})}{r^{2}}. \label{18}
\end{equation}

This metric is not asymptotically flat due to the
presence of a global monopole, by setting $f(r_{+})=0$, one can easely find the locations of the outer and inner horizons given by
\begin{equation}
r_{\pm}=M\pm\sqrt{M^{2}-Q_{h}^{2}}. \label{19}
\end{equation}

Note that when the global monopole is introduced, the AMD mass $M_{a}$ and AMD charge $Q_{h,a}$ of the global monopole black hole are different from the mass and charge parameter, $M$ and $Q$, respectively. In particular, the Komar's  integrals gives \cite{sharif} $M_{a}=(1-8\pi\eta^{2})M$, and $Q_{h,a}=(1-8\pi\eta^{2})Q_{h}$.  The electromagnetic four potential of the black hole is given by 
\begin{equation}
\tilde{A}_{\mu}=\left(-\frac{Q_{h}}{r},0,0,0\right). \label{20}
\end{equation}

Now we aim to introduce the effect of quantum gravity, therefore, we can start from the modified commutation relation \cite{kempf}, given as \cite{qgv1,qgv2,ali}
\begin{equation}
\Delta x \Delta p \geq \frac{\hbar}{2}\left(1+\alpha_{GUP} (\Delta p)^{2}\right). \label{21} 
\end{equation}

Note that $\alpha_{GUP}=\alpha_{0}/M_{p}=\alpha_{0}l_{p}^{2}/\hbar^{2}$, where $M_{p}$ is the Planck mass, $l_{p}$ is the Planck length, and $\alpha_{0}$ is a dimensionless parameter. With that in mind, one can show that, the position, momentum, energy and frequency
operators are modified. Under the effect of minimum length, we can now write the modified Dirac equation in curved spacetime for a particle with mass $m$ and equivalent charge $q_{h}$. Based on the same arguments as presented by \cite{qgv1,qgv2}, for the modified Dirac equation we can write 
\begin{eqnarray}
&&-\gamma^{t}\partial_{t}\Psi=\left(\gamma^{i}\partial_{i}+\gamma^{\mu}\Omega_{\mu}+\gamma^{\mu}\frac{i}{\hbar}q_{h} \tilde{A}_{\mu}+\frac{m}{\hbar}\right)\times \notag \\
&&\left(1+\alpha_{GUP} \hbar^{2}\partial_{j}\partial^{j}-\alpha_{GUP} m^{2} \right)\Psi. \label{22}
\end{eqnarray}%
 
The complete state of such particles can be described by a spinor field $\Psi$, which on the other hand can always be written as as a linear superposition of the spin-up and
spin-down states. However, if we perform a measurement, we always find the particle in one of these states. Let us consider the spin up case by choosing the following ansatz
\begin{equation}
\Psi _{\uparrow }\left( t,r,\theta ,\varphi\right) =\left(
\begin{array}{c}
A\left( t,r,\theta ,\varphi\right)  \\
0\\
B\left( t,r,\theta ,\varphi\right)\\
0
\end{array}
\right) e^{\left( \frac{i}{\hbar }I_{\uparrow }\left( t,r,\theta ,\varphi\right)\right)}.
 \label{23}
\end{equation}

Without loss of generality, by considering the metric \eqref{17}, we can select the following ansatz for the $\gamma^{\mu}$ matrices
\begin{eqnarray}\nonumber
\gamma ^{t}=\frac{1}{\sqrt{f(r)}}\left(
\begin{array}{cc}
i & 0 \\
0 & -i
\end{array}
\right), \,\,\,\, \gamma ^{r}=\sqrt{f(r)}\left(
\begin{array}{cc}
0 & \sigma ^{3} \\
\sigma ^{3} & 0
\end{array}
\right),             
\end{eqnarray}
\begin{eqnarray}\nonumber
\gamma ^{\theta}=\sqrt{g^{\theta\theta}}\left(
\begin{array}{cc}
0 & \sigma ^{1} \\
\sigma ^{1} & 0
\end{array}
\right), \,\,\,\, \gamma ^{\varphi}=\sqrt{g^{\varphi\varphi}}\left(
\begin{array}{cc}
0 & \sigma ^{2} \\
\sigma ^{2} & 0
\end{array}
 \right) ,            
\end{eqnarray}
where $\sigma ^{i } \,(i=1,2,3)$ are the Pauli matrices
\[
\sigma ^{1}=\left(
\begin{array}{cc}
0 & 1 \\
1 & 0
\end{array}
\right), \, \sigma ^{2}=\left(
\begin{array}{cc}
0 & -i \\
i &  0
\end{array}
\right), \,\sigma ^{3}=\left(
\begin{array}{cc}
1 & 0 \\
0 & -1
\end{array}
\right) .
\]

We can now apply the WKB approximation which consists in keeping only the contribution of leading terms of $\hbar$. Keeping in mind that and by substituting Eq. \eqref{23} into Eq. \eqref{22}, one finds:
\begin{equation}
0=-iA\frac{(\partial_{t}I_{\uparrow })}{\sqrt{f}}+iAm\Xi-B\sqrt{f}(\partial_{r}I_{\uparrow }) \Xi-iA\frac{q_{h}\tilde{A}_{t}}{\sqrt{f}}\Xi \label{24}
\end{equation}
\begin{equation}
0=iB\frac{(\partial_{t}I_{\uparrow })}{\sqrt{f}}+iBm\Xi-A\sqrt{f}(\partial_{r}I_{\uparrow }) \Xi+iB\frac{q_{h}\tilde{A}_{t}}{\sqrt{f}}\Xi \label{25}
\end{equation}
\begin{eqnarray}
0&=&A\Big[-(1-\alpha_{GUP} m^{2})\sqrt{g^{\theta\theta}} (\partial_{\theta}I_{\uparrow })\notag\\
&&+\alpha_{GUP} \sqrt{g^{\theta\theta}}(\partial_{\theta}I_{\uparrow })\Sigma-i\sqrt{g^{\varphi\varphi}}(\partial_{\varphi}I_{\uparrow })\Xi \Big]\label{26}\\
0&=&B\Big[-(1-\alpha_{GUP} m^{2})\sqrt{g^{\theta\theta}} (\partial_{\theta}I_{\uparrow })\notag\\
&&+\alpha_{GUP} \sqrt{g^{\theta\theta}}(\partial_{\theta}I_{\uparrow })\Sigma-i\sqrt{g^{\varphi\varphi}}(\partial_{\varphi}I_{\uparrow })\Xi \Big]\label{27}
\end{eqnarray}
where we have used
\begin{eqnarray}
\Xi&=&1-\alpha_{GUP} m^{2}-\alpha_{GUP} g^{rr}(\partial_{r}I_{\uparrow })^{2}\notag\\
&&-\alpha_{GUP} g^{\theta\theta}(\partial_{\theta}I_{\uparrow })^{2}-\alpha_{GUP} g^{\theta\theta}(\partial_{\theta}I_{\uparrow })^{2}\label{28}
\end{eqnarray}
and
\begin{equation}
\Sigma=g^{rr}(\partial_{r}I_{\uparrow })^{2}+g^{\theta\theta}(\partial_{\theta}I_{\uparrow })^{2}+g^{\varphi\varphi}(\partial_{\varphi}I_{\uparrow })^{2}. \label{29}
\end{equation}

A careful analysis shows that \cite{qgv1}, only the radial part remains to be discussed. The Killing vectors of the metric \eqref{17} which are dictated by the spacetime symmetries allows us to use the following ansatz for the action 
\begin{equation}
I_{\uparrow }(t,r)\equiv I_{0}(t,r)=-E_{a}t+R(r)+C. \label{30}
\end{equation}

In which $E_{a}$, is the energy of the particle in the presence of the global monopole given by  $E_{a}=(1-8\pi\eta^{2}) E$, and $C$ is some complex constant. In similar way, the equivalent charge of the particle $q_{h}$,  should be replaced by $q_{h,a}=(1-8\pi\eta^{2}) q_{h}$. Next, if we insert the action \eqref{30}, into Eqs. \eqref{24} and \eqref{25} and by canceling $A$ and $B$ gives
\begin{equation}
A_{6} (R^{\prime})^{6}+A_{4}(R^{\prime})^{4}+A_{2}(R^{\prime})^{2}+A_{0}=0, \label{31}
\end{equation}
in which
\begin{equation}
A_{6}= \alpha^{2}_{GUP}f^{4}
\end{equation}
\begin{equation}
A_{4}= \alpha_{GUP} f^{3}(3m^{2}\alpha_{GUP} -2)-\alpha^{2}_{GUP}f^{2}q_{h,a}^{2}\tilde{A}_{t}^{2}
\end{equation}
\begin{equation}
A_{2}= f^{2}\Big[(1-\alpha_{GUP} m^{2})^{2}-2\alpha_{GUP} m^{2}(1-\alpha_{GUP} m^{2})\Big]\notag
\end{equation}
\begin{equation}
+2\alpha_{GUP} f q_{h,a} \tilde{A}_{t}\Big[-E_{a}+q_{h,a}\tilde{A}_{t}(1-\alpha_{GUP} m^{2})\Big]
\end{equation}
\begin{equation}
A_{0}= m^{2}f(1-2m^{2})^{2}-\Big[-E_{a}+q_{h,a}\tilde{A}_{t}(1-\alpha_{GUP} m^{2})\Big]^{2}.
\end{equation}

We can now analyze and simplify the solution of the radial part by neglecting the higher order terms of $\alpha_{GUP}$, then, the solution for $R(r)$ is calculated as
\begin{equation}
R_{\pm}(r)=\pm \int \frac{\sqrt{\Delta^{2}-m^{2}f(1-2m^{2}\alpha_{GUP})}}{f\sqrt{1-2m^{2}\alpha_{GUP}}}dr\end{equation}
where 
\begin{equation}
\Delta=E_{a}-q_{h,a}\tilde{A}_{t}(1-m^{2}\alpha_{GUP}).\notag
\end{equation}

Let us now expand the function $ f(r)$ in Taylor's series near the horizon 
\begin{equation}
f(r_{+})\approx f^{\prime }(r_{+})(r-r_{+}),
\end{equation}%
and introduce the Feynman $i\epsilon$--prescription and make use of the formula $(r-r_{+}-i\epsilon)^{-1}=\mathcal{P}[(r-r_{+})^{-1}]+i\pi\delta(r-r_{+})$, where $\mathcal{P}$ denotes the principal part \cite{vanzo}, to get 
\begin{equation}
\text{Im}\,R_{\pm}(r_{+})=\pm  \frac{ \pi \Delta(r_{+})}{f^{\prime}(r_{+})\sqrt{1-2m^{2}\alpha_{GUP}}},
\end{equation}
or
\begin{equation}
\text{Im}\,R_{\pm}(r)=\pm  \frac{ \pi r_{+}^{2}(1-8\pi\eta^{2})}{r_{+}-r_{-}}\frac{E_{net}}{\sqrt{1-2m^{2}\alpha_{GUP}}}. \label{37}
\end{equation}

In the last equation we have used $E_{net}=E-q_{h}\tilde{A}_{t}(1-\alpha_{GUP} m^{2})$. As it was pointed out by \cite{Akhmedova1,Akhmedova2}, due to the temporal contribution the Hawking temperature turns out to be twice of the original temperature.  This difficulty, however, can easily be solved if we let the outside particle falls into the black hole with a $100\%$ chance of entering the black hole. It is evident now that, the corresponding probability of the ingoing particle should be 
\begin{equation*}
P_{-}\simeq \exp\left({-\frac{2}{\hbar}\text{Im}R_{-}}\right)=1,
\end{equation*}%
which also implies $\text{Im}\,I_{-}=\text{Im}\,R_{-}+\text{Im}\,C=0,$ therefore, $\text{Im}\,C=-\text{Im}\,R_{-}$. But for the outgoing particle we have, $
\text{Im}\,I_{+}=\text{Im}\,R_{+}+\text{Im}\,C$. We see that Eq. \eqref{37} also suggests $R_{+}=-R_{-}$, thus, the probability for the outgoing particle reads
\begin{equation}
P_{+}=\exp \left(-\frac{2}{\hbar}\text{Im}I_{+}\right)\simeq \exp\left(-\frac{4}{\hbar}\text{Im}R_{+}\right).
\end{equation}

In complete analogy, as in ordinary quantum mechanics, we can define the tunneling rate of the particles tunneling from inside to outside the horizon as a ratio of the last two equations

\begin{equation}
\Gamma =\frac{P_{+}}{P_{-}}\simeq \exp{(-\frac{4}{\hbar}\text{Im}R_{+})}.
\end{equation}

Applying this result and make use of the Eq. \eqref{37} we find
\begin{equation}
\Gamma_{GUP}=\exp\left[-\frac{4 \pi }{\hbar}\frac{ r_{+}^{2} (1-8\pi\eta^{2})}{r_{+}-r_{-}}\frac{E_{net}}{\sqrt{1-2m^{2}\alpha_{GUP}}}\right].
\end{equation}

In order to find the Hawking temperature we have to compere the last equation with the Boltzmann factor
$ \exp (- (1-8\pi\eta^{2}) E_{net}/T_{GUP})$. The Hawking temperature reads
\begin{eqnarray}
T_{GUP}&=&\frac{\hbar}{4 \pi}\frac{r_{+}-r_{-}}{r_{+}^{2}}\sqrt{1-2m^{2}\alpha_{GUP}}\notag\\
&=&T_{H}\sqrt{1-2m^{2}\alpha_{GUP}}.
\end{eqnarray}

Note that $T_{H}$ is the semi--classical Hawking temperature for the charged and magnetized black hole given by 
\begin{equation}
T_{H}=\frac{\hbar}{2\pi}\frac{\sqrt{M^{2}-Q_{e}^{2}-Q_{m}^{2}}}{\left(M+\sqrt{M^{2}-Q_{e}^{2}-Q_{m}^{2}} \right)^{2}}.
\end{equation}

We now see that by setting $\alpha_{GUP}=0$ the last two equations coincides. On the other hand, the GUP parameter $\alpha_{GUP}$, slows down the increase of the Hawking temperature caused by the evaporation process, which means that GUP may prevent a black hole from complete evaporation and leave remnants.

\section{Temperature and Entropy correction Beyond Semiclassical Approximation}

Now we would like to consider the quantum effect on the Hawking temperature, therefore, we may write the action as  following \cite{banerjee}
\begin{equation}
I(t,r)=I_{0}(t,r)+\sum_{i}\hbar^{i}I_{i},
\end{equation}

But, since $S_{0}$ has the dimension of $\hbar$, the proportionality constants should have the dimension of inverse of $\hbar^{i}$. Again in the units $G = c = k_{B} = 1$ the Planck constant $\hbar$ is of the order of square of the Planck Mass $M_{p}$ and so from dimensional analysis the proportionality constants have the dimension of $M^{2i}$ where $M$ is the mass of black hole \cite{banerjee,akbar}. However, as is noted in the Ref. \cite{akbar}, for the
sake of simplicity, in the above units, the Planck constant is also
of the order of square of the Planck Length $l_{p}$, therefore the proportionality
constants have also the dimension of $r_{+}^{2}$
\begin{eqnarray}
I(t,r)&=&I_{0}(t,r)+\sum_{i}\beta_{i}\frac{\hbar^{i}}{r^{2i}}I_{0}(r,t)\\
&=&\left(1+\sum_{i}\beta_{i}\frac{\hbar^{i}}{r^{2i}}\right) I_{0}(r,t).
\end{eqnarray}

Let us now recall that the general GUP can be expressed as follows \cite{brito1,brito2}
\begin{equation}
\Delta x\Delta p_{GUP}\geq \hbar \left( 1-\frac{y}{\hbar }\Delta p_{GUP}+%
\frac{y^{2}}{\hbar ^{2}}(\Delta p_{GUP})^{2}\right) ,  \label{26n}
\end{equation}
in which $y=\alpha _{GUP}l_{p}$ and $\alpha _{GUP}$ is a dimensionless
positive constant. This equation can also can be rewritten as 
\begin{equation}
\Delta p_{GUP}\geq \frac{\hbar (\Delta x+y)}{2y^{2}}\left( 1-\sqrt{1-\frac{%
4y^{2}}{(\Delta x+y)^{2}}}\right) ,  \label{27n}
\end{equation}

From now on we will set the also the Planck constant to unity i.e. $\hbar =1$, therefore we have $l_{p}=G=c=\hbar =k_{B}=1 $. If we expand the last equation in Taylor series we get
\begin{equation}
\Delta p_{GUP}\geq \frac{1}{\Delta x}\left[ 1-\frac{\alpha _{GUP}}{2\Delta x}%
+\frac{\alpha _{GUP}^{2}}{2(\Delta x)^{2}}+\cdots \right] .  \label{28n}
\end{equation}

Next, we can now make use of the uncertainty
principle and its saturated form \cite{brito1,brito2,soj} $\Delta x\Delta p\geq 1 $
and $E_{a} \Delta x\geq 1 $, to find 
\begin{equation}
E_{a,GUP}\geq E_{a} \left[ 1-\frac{\alpha _{GUP}}{2(\Delta x)}+\frac{\alpha
_{GUP}^{2}}{2(\Delta x)^{2}}+\cdots \right] .  \label{31n}
\end{equation}%

Where $E_{a,GUP}$ is the quantum corrected energy of the particle. We can now apply the Hamilton--Jacobi method and study the tunneling probability of a particle with corrected energy to obtain 
\begin{equation}
\Gamma_{GUP}=\exp\left(-2 \text{Im} I_{GUP}\right)=\exp\left(- \frac{4 \pi E_{a,GUP}}{f^{\prime}(r_{+})}\right).
\end{equation}

Comparing this result with the Boltzmann factor $\exp({-E_{a}/T_{GUP}})$, we derive the GUP corrected temperature
\begin{equation}
T_{GUP}=T_{H}\left[ 1-\frac{\alpha _{GUP}}{4 r_{+}}+\frac{\alpha
_{GUP}^{2}}{8 r_{+}^{2}}+\cdots \right] ^{-1}.  \label{33n}
\end{equation}

Note that in the last equation we have chosen $\Delta x=2r_{+}$. The modified probability of the outgoing particle thus can be written
as 
\begin{equation}
\tilde{\Gamma}_{GUP}=\exp\Big[-\left(1+\sum_{i}\frac{\beta_{i}}{r_{+}^{2i}}\right) \frac{4 \pi E_{a,GUP}}{f^{\prime}(r_{+})}\Big].
\end{equation}

Using the same arguments as in the last section, by comparing
with the Boltzmann factor $\exp({-E_{a}/\tilde{T}_{GUP}})$ one can easily get the Hawking temperature
\begin{equation}
\tilde{T}_{GUP}=\frac{T_{H}}{\left(1+\sum_{i}\frac{\beta_{i}}{r_{+}^{2i}}\right)\left[ 1-\frac{\alpha _{GUP}}{4 r_{+}}+\frac{\alpha
_{GUP}^{2}}{8 r_{+}^{2}}+\cdots \right]}.
\end{equation}

Let us now calculate the entropy corrections. To do so, we need to write the first law of black hole mechanics for the Reissner--Nordstr\"{o}m black hole with a global monopole given by
\begin{equation}
\tilde{S}_{GUP}=\int \frac{1}{\tilde{T}_{GUP}}\left(dM_{a}+\frac{Q_{e}}{r_{+}}dQ_{e,a}+\frac{Q_{m}}{r_{+}}dQ_{m,a}\right).
\end{equation}

We can solve the above integral by considering first the differential of the event horizon
equation $r_{+}$, to get
\begin{equation}
\frac{r_{+}-M}{r_{+}}dr_{+}=dM+\frac{Q_{e}}{r_{+}}dQ_{e}+\frac{Q_{m}}{r_{+}}dQ_{m}.
\end{equation}

In this way, combining the last two equations and use the following relations $M_{a}=aM$ and $Q_{h,a}=aQ_{h}$, we end up with the following integral
\begin{eqnarray}\notag
\tilde{S}_{GUP}&=& 4\pi a\int \left(1+\sum_{i}\frac{\beta_{i}}{r_{+}^{2i}}\right)\frac{r_{+}\left(r_{+}-M\right)}{r_{+}-r_{-}} \\\notag
&\times & \left[ 1-\frac{\alpha _{GUP}}{4 r_{+}}+\frac{\alpha
_{GUP}^{2}}{8 r_{+}^{2}}+\cdots \right]  dr_{+}\\\notag
&=& 2\pi a\int \left(1+\sum_{i}\frac{\beta_{i}}{r_{+}^{2i}}\right)\\
&\times & \left[ 1-\frac{\alpha _{GUP}}{4 r_{+}}+\frac{\alpha
_{GUP}^{2}}{8 r_{+}^{2}}+\cdots \right] r_{+}dr_{+}.
\end{eqnarray}

Note that in the second line we have used the fact that $r_{+}-r_{-}=2\sqrt{M^{2}-Q_{h}^{2}}$ and $r_{+}-M=\sqrt{M^{2}-Q_{h}^{2}}$. Finally, the last integral can be easily evaluated to find the expression for the logarithmic corrected GUP entropy
\begin{eqnarray}\notag
\tilde{S}_{GUP} &=&\pi r_{+}^{2}a-\frac{\alpha_{GUP} \pi a \ln r_{+}}{2}-\frac{\alpha_{GUP}^{2} \pi a}{4 r_{+}}\\\notag
&+&\beta_{1}\pi a \left(2 \ln r_{+}+\frac{\alpha_{GUP}}{2 r_{+}}-\frac{\alpha_{GUP}^{2}}{8 r_{+}^{2}}\right)\\
&+& \beta_{2} \pi a \left( - \frac{1}{r_{+}^{2}}+\frac{\alpha_{GUP}}{6 r_{+}^{3}}-\frac{\alpha_{GUP}^{2}}{16 r_{+}^{4}}\right)+\cdots
\end{eqnarray}

In terms of the Bekenstein--Hawking entropy the GUP corrected entropy can be written as
\begin{eqnarray}\notag
\tilde{S}_{GUP} &=&S_{BH}-\frac{\alpha_{GUP} \pi a \ln S_{BH}}{4}-\frac{\alpha_{GUP}^{2}\pi^{3/2}a^{3/2}}{4 \sqrt{S_{BH}}}\\\notag
&+&\beta_{1}\pi a \ln S_{BH}+\frac{\beta_{1}\pi ^{3/2}a^{3/2}\alpha_{GUP}}{2 \sqrt{S_{BH}}}\\\notag
&-&\frac{\beta_{1}\alpha_{GUP}^{2} \pi^{2}a^{2}}{8 S_{BH}}-\frac{\beta_{2}\pi^{2}a^{2}}{S_{BH}}+\frac{\beta_{2}\alpha_{GUP} \pi^{5/2}a^{5/2}}{6 S^{3/2}_{BH}}\\
&-&\frac{\beta_{2}\alpha_{GUP}^{2}\pi^{3}a^{3}}{16 S_{BH}^{2}}+Const+\cdots
\end{eqnarray}

In which the first term $S_{BH}$ is the Bekenstein--Hawking entropy given by $S_{BH}=\pi r_{+}^{2}a $ followed by the logarithmic correction terms. It's interesting to see that, the quantum corrected GUP entropy is shown to depend on the global monopole parameter $\eta^{2}$, GUP parameter $\alpha_{GUP}$, and quantum correction coefficients.  Now if we assume that the form of the GUP corrected specific heat capacity at constant charge remain the same, and for simplicity, we will consider here only the quantum gravity effects by setting $\beta_{1}=\beta_{2}=0$. We find
\begin{eqnarray}\notag
C_{GUP}&=&T_{GUP}\frac{\partial S_{GUP}}{\partial T_{GUP}}\\
&=& T_{GUP} \frac{\partial S_{GUP}}{\partial r_{+}} \left(\frac{\partial T_{GUP}}{\partial r_{+}}\right)^{-1}.
\end{eqnarray}

Using the above definition we find the following result for the GUP specific heat capacity 
\begin{eqnarray}\notag
C_{GUP}&=& \frac{a\pi(r_{+}-r_{-}) (\alpha_{GUP}^{2}-2\alpha_{GUP}r_{+}+8r_{+}^{2})}{4\left[(\alpha_{GUP}^{2}-2\alpha_{GUP}r_{+})r_{+}^{5}+16r_{-}r_{+}^{6}-8r_{+}^{7}\right]}\\
&\times& \left(\alpha_{GUP}^{2}r_{+}^{3}-2\alpha_{GUP} r_{+}^{4}+8r_{+}^{6}\right).
\end{eqnarray}

\begin{figure}
\centering
\begin{minipage}{.45\textwidth}
\centering
\includegraphics[width=.97\linewidth]{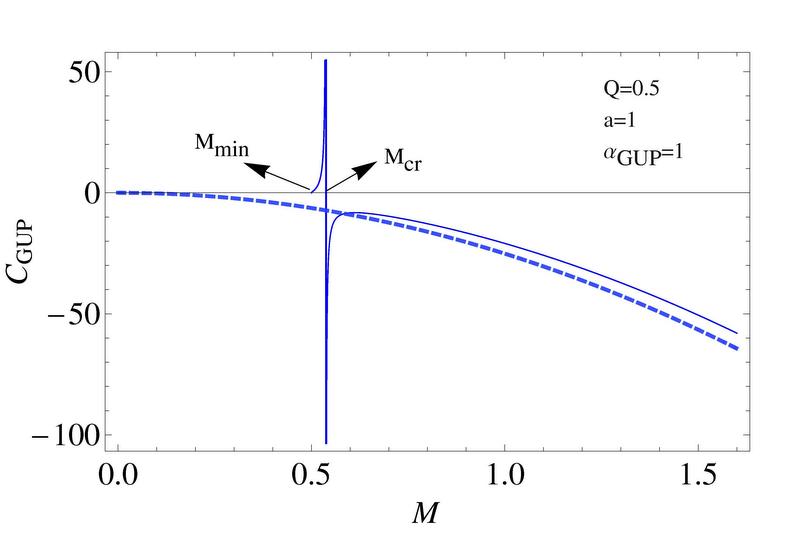}
\caption{\small \textit{ Plots for the GUP specific heat capacity corresponding to $Q=0.5$ and $a=\alpha_{GUP}=1$. The dashed curve gives the standard specific heat capacity for $Q=\alpha_{GUP}=0$.}}
\end{minipage}
\end{figure}

As expected, from Fig.1 we can observe that the standard specific heat capacity (given by the dashed line) goes to zero when $M\to 0$.  On the other hand, the GUP corrected specific heat has a vertical asymptote at some critical mass $M = M_{cr}$, which shows that a thermodynamic phase transition happened from $C_{GUP}<0$ (unstable phase) to $C_{GUP} > 0$ (stable phase). Morover the GUP corrected specific heat goes to zero at some minimal mass $M_{min}$, which indicates that the black hole cannot exchange radiation with the surrounding space \cite{khalil,feng}. Therefore, the GUP effects may prevent a black hole from complete evaporation and leave remnants.  We also see that the GUP corrected specific heat capacity depend on the global monopole parameter.

\section{Conclusion}

In this paper, we have successfully recovered the GUP corrected Hawking temperature of Dirac magnetic monopoles under the generalized uncertainty relation by solving the modified Dirac equation applied to the magnetized particles in the spacetime background with a global monopole. We have used the WKB approximation and the separation of variables and have take into account that the ADM mass/charge of the black hole shifts by a factor of $1-8\pi\eta^{2}$ from the black hole mass and black hole charge parameters. We have successfully recovered the logarithmic GUP entropy corrections and GUP specific heat capacity beyond the semiclassical approximation. It is argued that, the GUP corrected specific heat has a vertical asymptote at some critical mass $M = M_{cr}$ and goes to zero at some minimal mass $M_{min}$. This indicates that the black hole cannot exchange radiation with the surrounding space \cite{khalil,feng} which may prevent a black hole from complete evaporation and leave remnants.

\end{document}